\documentstyle[aps,prl,preprint,floats,epsf]{revtex}

\begin{document}

\preprint{\tighten\vbox{\hbox{\hfil CLNS 97/1519}
                        \hbox{\hfil CLEO 97-26}
}}

\title{Study of Semileptonic Decays of $B$ Mesons to Charmed Baryons }

\maketitle
\tighten

\begin{abstract}
Using data collected by the CLEO II detector at a center-of-mass energy on 
or near the $\Upsilon(4S)$ resonance, 
we have determined the 90\% confidence level upper limit    
${\cal B}(\overline{B}\rightarrow\Lambda^+_ce^-X )/
{\cal B}(\overline{B}\rightarrow(\Lambda^+_c~{\rm or}~
\overline{\Lambda}^-_c)X)<0.05$ for electrons with 
momentum above 0.6 GeV/$c$.  We have also obtained the limit  
${\cal B}(B^-\rightarrow\Lambda^+_c\overline{p}e^-\overline{\nu}_e)/
{\cal B}(\overline{B}\rightarrow\Lambda^+_c\overline{p} X)<0.04$ 
at the 90\% confidence level and measured the ratio
 ${\cal B}(\overline{B}\rightarrow\Lambda^+_c\overline{p}X)/
{\cal B}(\overline{B}\rightarrow(\Lambda^+_c~{\rm or}~
\overline{\Lambda}^-_c)X)=0.57\pm 0.05\pm 0.05$. 
\end{abstract}
\newpage
{
\renewcommand{\thefootnote}{\fnsymbol{footnote}}

\begin{center}
G.~Bonvicini,$^{1}$ D.~Cinabro,$^{1}$ R.~Greene,$^{1}$
L.~P.~Perera,$^{1}$ G.~J.~Zhou,$^{1}$
M.~Chadha,$^{2}$ S.~Chan,$^{2}$ G.~Eigen,$^{2}$
J.~S.~Miller,$^{2}$ C.~O'Grady,$^{2}$ M.~Schmidtler,$^{2}$
% ...
\end{center}
 
\small
\begin{center}
$^{1}${Wayne State University, Detroit, Michigan 48202}\\
$^{2}${California Institute of Technology, Pasadena, California 91125}\\
% ...
\end{center}

%

\medskip
\centerline{\bf INTRODUCTION}
\medskip
In the naive spectator model, most $B$ mesons decay through the spectator 
diagram with semileptonic decays occurring by ``external'' $W$-emission: 
$b\rightarrow cW;~W\rightarrow\ell\overline{\nu}_{\ell}$.  
In this picture, charmed baryon production occurs when two quark-antiquark 
pairs from the vacuum bind with the charm quark and the spectator antiquark to 
form a $\Lambda^+_c(cud)$ plus an antinucleon $\overline{N}$.  In this paper 
we attempt to isolate the magnitude of this external $W$-emission 
spectator diagram in charmed baryon decays by measuring 
$\overline{B}\rightarrow\Lambda^+_c e^-X$ and 
$B^-\rightarrow\Lambda^+_c\overline{p}e^-\overline{\nu}_e$.  
For normalization modes, we also measure 
$\overline{B}\rightarrow\Lambda^+_c\overline{p}X$ and  
$\overline{B}\rightarrow(\Lambda_c^+~{\rm or}~\overline{\Lambda}^-_c)X$.  
Throughout this paper charged conjugate modes are implicit.   

If $B\rightarrow baryons$ does indeed occur through external $W$-emission as 
outlined above,  then the decay 
$\overline{B}\rightarrow\Lambda^+_c\overline{N}Xe^-\overline{\nu}_{\ell}$
will occur~\cite{ref3}.  We can estimate the magnitude of 
$R={\cal B}(\overline{B}\rightarrow\Lambda^+_c\overline{N}e^-
\overline{\nu}_{\ell})
/{\cal B}(\overline{B}\rightarrow\Lambda^+_c\overline{N}X)$ by using the
naive expectation for the semileptonic branching ratio in these decays.
The ($\overline{c} s$) and $(\tau \overline{\nu}_{\tau})$ 
contributions are absent due to
the limited available phase space so a maximum of 20\% is expected 
for the ratio $R$.  Alternately, one might anticipate that
${\cal B}(\overline{B}\rightarrow\Lambda^+_cXe^-\overline{\nu}_e)/
{\cal B}(\overline{B}\rightarrow
\Lambda^+_cX)$ is comparable to the measurements of 
${\cal B}(\overline{B}\rightarrow DXe^-\overline{\nu}_e)/
{\cal B}(\overline{B}\rightarrow DX)\simeq 12\%$~\cite{ref5}.  

There are two other baryon production mechanisms in $B$ decay, neither making 
a contribution to semileptonic decay.  
In one, the $W$ is emitted internally and decays to $(\overline{c} s)$, 
leading to $\Xi_c\overline{\Lambda}_c$ final states.    
This mechanism was studied in  
a previous CLEO paper, which looked at the charge correlations 
between $\Lambda_c$'s and leptons from $B$ decay and found 
$R_{\Lambda_c}=N_{\overline{\Lambda}_c^-\ell^+}/N_{\Lambda_c^+\ell^+}=
{\cal B}(\overline{B}\rightarrow\overline{\Lambda}_c^-X) 
{\cal B}(B\rightarrow X\ell^+\nu_{\ell})/
{\cal B}(\overline{B}\rightarrow\Lambda^+_cX) 
{\cal B}(B\rightarrow X\ell^+\nu_{\ell})=
0.19\pm 0.13\pm 0.04$ which is directly related to  
${\cal B}(b\rightarrow c\overline{c}s)/
{\cal B}(b\rightarrow c\overline{u}d)$~\cite{ref4}.  
For $\Lambda_c X$ final states, we cannot rule out the possibility 
in our analysis that we are observing decays of the type 
$\overline{B}\rightarrow\Xi_c\overline{\Lambda}_c$, as we cannot tag the 
parent $B$ meson in the $\overline{B}\rightarrow\Lambda^+_cX$ analysis.  
Therefore, the yields for this mode will be quoted as 
decays of the type $\overline{B}\rightarrow
(\Lambda^+_c~{\rm or}~\overline{\Lambda}^-_c)X$.  
Another mechanism is the internal emission of a $W$ followed by its 
decay to $(\overline{u}d)$.  
Measurements of $B$ mesons decaying hadronically to 
charmed baryons indicate that this internal $W$ 
emission diagram may contribute 
significantly~\cite{ref20}.  A substantial contribution from this 
diagram would reduce the semileptonic decay width. 

The semileptonic branching ratio of $B$ mesons is
known to have a lower value than theoretical predictions~\cite{ref1}.  
These predictions assume a large 
external $W$-emission contribution in baryon 
decays.  The suggestion has been made that theory may underestimate
the $B$-hadronic width by neglecting $B$ decay channels to 
baryon states~\cite{ref2}.    If this is the case, hadronic
decays to charmed baryons could explain 
the low inclusive semileptonic branching ratio.  
The measurement of semileptonic
decays of $B$ mesons to $\Lambda_c$ will provide vital information 
on baryon production in $B$ decays. 
 
\bigskip
\centerline{\bf  Data sample and Event selection }
\bigskip

The data were taken with the CLEO II 
detector~\cite{ref11} at the Cornell Electron Storage Ring (CESR), and 
consist of 3.2 fb$^{-1}$ on the $\Upsilon(4S)$ resonance and 
1.6 fb$^{-1}$ at a center-of-mass energy 60 MeV below the resonance.  The 
on-resonance sample contains 3.4 million $B\overline{B}$ events and 
10 million continuum events.  We select hadronic events containing at 
least 4 charged tracks.  To suppress continuum background, we require the 
ratio of Fox-Wolfram 
moments~\cite{ref13} $R_2= H_2/H_0$ to satisfy $R_2\leq 0.35$.  
We reconstruct $\Lambda_c$'s in the $pK\pi$ decay mode.  
For the hadronic particle identification, a probability cut for
each target hadron is made which uses
information obtained from $dE/dx$ and time-of-flight detectors.  
For particle consistency, the probability cuts are chosen to be greater than:
0.0027 (within three standard deviations of the expected value) for pions, 
0.0001 for kaons, and  0.0003 for protons.
Continuum data are used to directly subtract backgrounds from non-$B\overline{B}$ 
events.  

Tagged signal Monte Carlo simulated events were used to obtain the 
signal efficiencies while 
$B\overline{B}$ Monte Carlo simulated events, with the 
signal channel removed, were used to 
estimate the background from $B$ decays to non-signal modes.  
The CLEO $B\overline{B}$ Monte-Carlo simulation generates baryonic decays
with a phenomenological model which is tuned to match the observed 
$\Lambda_c$ momentum spectrum.  We use $B\overline{B}$ 
Monte Carlo events where we force the 
$\overline{B}\rightarrow\Lambda^+_c X$, $\Lambda^+_c\rightarrow pK^-\pi^+$ 
decay chain to determine a detection efficiency of $0.36\pm 0.01$.

The $pK^-\pi^+$ invariant mass distributions are measured 
separately for the resonance and continuum data.  
The resonance data are fitted to a double Gaussian signal atop
a low-order polynomial background. 
In these fits, the width of the Gaussian
$\Lambda_c$ signal function is constrained 
to the value derived from the Monte Carlo
simulation.   After subtracting non-BBbar contributions using
off-resonance data scaled for luminosity and cross-section, 
we obtain a total sample of  
$4879\pm 296$ $\overline{B}\rightarrow
(\Lambda^+_c~{\rm or}~\overline{\Lambda}^-_c)X$ events from data.   
After scaling by the efficiency, we find a yield of $13552\pm 822\pm 802$ 
events, where the second (systematic) error includes contributions 
from the efficiency correction.    

\bigskip
\centerline{\bf  Study of  $\overline{B}\rightarrow\Lambda^+_ce^-X $ }
\bigskip

Due to the soft lepton momentum spectrum from this decay and the 
limited reconstruction efficiency for low momentum muons, we 
use only electrons in our analysis.  Electron identification  
relies on $E/p$ measurements derived from
the calorimeter and drift chamber, as well as specific ionization loss
measurements from the drift chamber.  
The requirement of $\ln(P_e/P_{\not e}) >3.0$ is imposed, where
$P_e(P_{\not e})$ is the probability that a given charged track 
is an electron (not an electron).
We choose a minimum momentum cutoff of 0.6 GeV/$c$ for 
these electrons to limit fake and secondary electron background sources.   
The maximum possible electron momentum for this decay is 1.5 GeV/$c$.  
Electron candidates are restricted to the polar angular region 
$\vert\cos\theta\vert\leq 0.71$.  We pair all $pK^-\pi^+$ candidates, 
selected as described above, with
additional tracks in the events passing the lepton identification 
requirement.  We then fit the $pK^-\pi^+$ invariant mass 
distributions on and off resonance for combinations passing these cuts. 

Figure\ \ref{fig1}(a) shows the fit to the $pK^-\pi^+$ 
invariant mass distribution for events that satisfy the above selection criteria.  
The resonance data (points) are fit to a double 
Gaussian signal over a second order  
polynomial background. The signal shape is fixed to that from the 
data in the $\overline{B}\rightarrow\Lambda^+_cX$ analysis.  
A similar fit has been performed on the $pK^-\pi^+$
invariant mass distribution from the continuum data (shown by the scaled 
histogram in the figure). The yields are given
in Table I.  

Besides continuum $\Lambda_c$'s, other 
sources of background are fake leptons and uncorrelated $\Lambda^+_c-e^-$ 
pairs.  
The number of fake leptons is obtained by running the same analysis, 
but using an electron anti-identification criterion: 
$\ln(P_e/P_{\not e}) <0$.  
The $pK^-\pi^+$ invariant mass is refit and this yield is 
scaled by the measured lepton misidentification probabilities.  
The uncorrelated background includes combinations where the 
$\Lambda^+_c$ originates from $\overline{B}$ decay and the lepton 
originates from $B$ decay or from a $\overline{B}$ if from an event 
where mixing took place.
This  background is estimated using  
$\overline{B}\rightarrow\Lambda^+_cX$ Monte Carlo events and examining decays where the $\Lambda^+_c$ and $e^-$ have opposite charges, but do 
not originate from the signal mode.  We check this procedure by comparing the 
data and Monte Carlo results obtained using $\Lambda^+_ce^+$ 
(wrong sign) combinations.  Wrong sign combinations will include primary 
leptons from one $B$ paired with $\Lambda^+_c$'s from the other $B$.  
We find consistency between the data and Monte Carlo wrong sign yields.  
The background predictions are given in Table I.      

The lepton minimum momentum cut of $0.6$ GeV/$c$ 
results in a model dependence.  
Larger multiplicity final states will have a lower efficiency due to the 
minimum momentum cut.  We find the efficiency using 
$B^-\rightarrow\Lambda_c^+\overline{p}e^-\overline{\nu}_e$ 
Monte Carlo events where 
the $B^+$ decays generically.  This efficiency for 
$B^-\rightarrow\Lambda_c^+\overline{p}e^-\overline{\nu}_e$, 
where the electron is prompt from the $B$ decay, is found to be 
17\%.  Monte Carlo events from the chain 
$\overline{B}^0\rightarrow\Lambda_c^+\overline{\Delta}^0e^-\overline{\nu}_e$, 
$\overline{\Delta}^0\rightarrow\overline{p}\pi^+$ 
were also generated to measure the 
efficiency.  This mode adds one extra pion to the total decay chain although 
more could be present in other decays such as 
$\overline{B}\rightarrow\Sigma_c\overline{\Delta} e\nu$.  
Differences between efficiencies 
for $B^-$ and $\overline{B}^0$  are found to be negligible.  After all 
other cuts, we find that 73\% of the events from 
$B^-\rightarrow\Lambda^+_c\overline{p}e^-\overline{\nu}_e$ 
pass our electron momentum cut while only 45\% of the events from  
$\overline{B}^0\rightarrow\Lambda_c^+
\overline{\Delta}^0e^-\overline{\nu}_e$ pass.  Because the total 
efficiency is dependent on the number of pions in the final state, 
we choose to quote a partial branching fraction where the lepton 
momentum is greater 
than 0.6 GeV/$c$.  In this electron momentum range, the efficiency for  
$\overline{B}\rightarrow\Lambda^+_c\overline{p}e^-\overline{\nu}_e$ 
is $0.239\pm 0.005$ which is consistent with the efficiency for other 
modes with extra pions.  In addition to assigning a systematic error 
due to efficiency determination, we add in quadrature errors from 
the fake lepton and uncorrelated background source estimates to 
obtain the total systematic error.  

\bigskip
\centerline{\bf  Search for $B^-\rightarrow 
\Lambda^+_c\overline{p}e^-\overline{\nu}_e$ }
\bigskip

The signature of $B^-\rightarrow \Lambda^+_c\overline{p}e^-\overline{\nu}_e$ 
is a baryon-lepton-antiproton
combination which has a recoil mass consistent with that of a neutrino, 
approximating the $B$ momentum as zero.  
Candidate $\Lambda^+_c$'s, electrons, and antiprotons for the 
analysis must satisfy requirements similar to those discussed above.  
We then require that the approximation of the squared mass of the 
neutrino,   
$\widetilde{M}_\nu^2\equiv (E_{\mbox{beam}}-E_{\Lambda_c} - 
E_e)^2-({\bf{p}}_{\Lambda_c}+{\bf{p}}_e)^2$ be greater than -2 
(GeV/c$^2$)$^2$.  
In addition, we place an angular cut of $\cos\theta_{\Lambda_c-e} <-0.2$, 
where $\theta_{\Lambda_c-e}$ is the angle  
between the $\Lambda^+_c$ and electron.  
In Figure\ \ref{fig1}(b) we show the 
$\Lambda^+_c$ invariant mass distribution for combinations passing all of 
these cuts.  This distribution is fit as before; results are given
in Table~I.

Backgrounds to this process stem from three sources: fake antiprotons or 
electrons, non-$B\overline{B}$ events, and secondary electrons or 
antiprotons.  Fake antiprotons and electrons are considered separately.  
We use the same methods as described above to determine each contribution.        
The continuum background is measured using the off-resonance data scaled 
for luminosity and cross section.  The remaining background events, in which electrons 
come from
the decay chain $\overline{b}\rightarrow\overline{c}\rightarrow
\overline{s}e\nu$, 
can be estimated using Monte Carlo 
simulation.  The 
wrong sign data and Monte Carlo results are compared and again found 
to agree well.

We find efficiency using the   
$B^-\rightarrow\Lambda_c^+\overline{p}e^-\overline{\nu}_e$ 
Monte Carlo events where 
the $B^+$ decays generically.  The efficiency is found to be 
$0.094\pm 0.003$.  Systematic errors are assigned for each of the 
background source estimates 
and the efficiency determination as described above.

\bigskip
\centerline{\bf  Study of $\overline{B}\rightarrow\Lambda^+_c\overline{p}X$  }
\bigskip

We pair all $\Lambda^+_c$ and $\overline{p}$ candidates using the 
$\Lambda^+_c$ selection as described above.  
For the $\overline{p}$, in addition
to the cut on the proton probability of greater than 0.0003, we employ 
additional veto cuts on the particle identification of 2$\sigma$ for 
$\pi$, $K$, and electron to reduce fake antiprotons.  We then
fit the $\Lambda^+_c$ invariant mass.  The observed  
$\Lambda^+_c$ signal area then
measures the number of  $\Lambda^+_c$-$\overline{p}$ correlations.
Figure\ \ref{fig2} shows the fit to the data.

We are looking for decays where the 
$\Lambda^+_c$ and $\overline{p}$ have opposite
charge and both are primary from the $B$ decay.  
Backgrounds are categorized into two sources: 
secondary $\overline{p}$ (not primary from a $B$ decay), 
and fake $\overline{p}$.  
The first background source is estimated by using 
$\overline{B}\rightarrow\Lambda^+_cX $ Monte Carlo
events as above.    Once again we check this 
procedure by comparing wrong sign data yields to our Monte Carlo wrong sign 
prediction.  Proton misidentification probabilities
are also measured directly from the data by using a pion sample 
from $ K^0_S\rightarrow\pi^+\pi^-$ where $ K^0_S$'s are selected by
a secondary vertex finder.  After applying the veto cuts for 
kaons, pions, and electrons, the fake probability derived from just using 
the pion rate is found to be consistent with the 
species averaged rate.  The background 
contribution of fake $\overline{p}$'s is obtained by running 
the same analysis 
without the particle identification cuts for the $\overline{p} $ 
correlated with $\Lambda^+_c$.  The yields obtained from fits to the 
$pK^-\pi^+$ mass are then multiplied by the faking 
$\overline{p}$ probabilities
and weighted by momentum.  This procedure will yield an upper limit on the 
number of fakes, as real proton tracks are double counted in our procedure.    
We studied the overcounting rate and assign a systematic error based on the 
difference between the number of protons counted with no identification 
criteria and the number counted using the anti-identification criteria.

We have also measured the absolute proton identification efficiency as 
a function of momentum for the cuts in this analysis using a sample 
of $\Lambda\rightarrow p\pi^-$ events.  
The differences in the identification efficiencies between 
data and Monte Carlo are then
used to calculate the systematic error in the 
${\cal B}(\overline{B}\rightarrow\Lambda^+_c\overline{p}X)/
{\cal B}(\overline{B}\rightarrow\Lambda^+_cX)$
measurement (Table\ \ref{tab1}).  The total systematic error for this 
mode is then derived from errors in the efficiency as well 
as background determinations.

\bigskip
\centerline{\bf SUMMARY }
\bigskip
Table\ \ref{tab1} 
summarizes the final numbers of candidates for the three signal modes.
After background subtraction, the number of wrong sign candidates 
observed in data are consistent with
our expectations based on Monte Carlo studies. 

We measure the ratio:
\begin{equation}
 {{\cal B}(\overline{B}\rightarrow\Lambda^+_c\overline{p}X)\over 
{{\cal B}(\overline{B}\rightarrow(\Lambda^+_c~{\rm or}~
\overline{\Lambda}^-_c)X)}}=
{(7669\pm 623\pm 385 )\over (13552\pm822\pm802)} = {0.57\pm 0.05\pm 0.05},
\end{equation}
consistent with the naively expected value of 50\%.

For the electron channels, the number of signal candidates 
is fully consistent with the 
expected background level, so we derive a 90\% confidence level 
upper limit for the ratio R, for $p_e\geq 0.6$ GeV/$c$: 
\begin{equation}
R={{\cal B}(\overline{B}\rightarrow\Lambda^+_c e^-X)\over 
{\cal B}(\overline{B}\rightarrow(\Lambda^+_c~{\rm or}~
\overline{\Lambda}^-_c)X)}={(259\pm196\pm143)\over(13552\pm822\pm802)} 
{<0.05}~{\rm at}~90\%~{\rm c.l.}
\end{equation}
If one assumes that all of the semileptonic decays proceed via the 
channel $B^-\rightarrow\Lambda_c^+\overline{p}e^-\overline{\nu}_e$, the 
upper limit on R would be 0.07 for the entire electron momentum range.  
Similarly, if all of the semileptonic decays were 
$\overline{B}^0\rightarrow\Lambda_c^+
\overline{\Delta}^0e^-\overline{\nu}_e$, the 
limit on R would be 0.11.  Our result is consistent with the derived limit 
based on a previous measurement where the charmed baryon is not observed 
and the lepton spectrum is extrapolated from a model of 
${\cal B}(\overline{B}\rightarrow X\overline{p}e^-\overline{\nu}_e) <0.16\% 
~@~90\%~$c.l.~\cite{ref9}.  This implies a limit on  
${\cal B}(\overline{B}\rightarrow\Lambda^+_ce^-X)
/{\cal B}(\overline{B}\rightarrow(\Lambda^+_c~{\rm or}~
\overline{\Lambda}^-_c)X) <5\% ~@~
90\%~$c.l.    

For the $\Lambda_c\overline{p}e\overline{\nu}_e$ channel, we find:
\begin{equation}
{{\cal B}(B^-\rightarrow\Lambda^+_c\overline{p}e^-\overline{\nu}_e)\over
{\cal B}(\overline{B}\rightarrow\Lambda^+_c\overline{p}X)}
{<0.04}~{\rm at}~90\%~{\rm c.l.}
\end{equation}
for the entire electron momentum range. 

Our limits on the semileptonic branching ratios do not support 
the hypothesis that 
the external $W$-emission diagram saturates charmed baryon production in 
$B$ decays.   While the ${\cal B}(\overline{B}\rightarrow\Lambda^+_ce^-X)$ 
measurement is limited by our knowledge of 
the possible decay states, the exclusive limit constrains the expected 
dominant mode below the corresponding rate measured for 
$B$ decays to charmed mesons.   The semileptonic decay rate 
from $B$ to baryons doesn't
add a large contribution to the total semileptonic $B$ decay rate if these semileptonics 
decays are dominated by modes of the type  
$\overline{B}\rightarrow\Lambda_c^+\overline{N}e^-\overline{\nu}_e$.  

\bigskip
\centerline{\bf ACKNOWLEDGEMENTS}
\smallskip
We gratefully acknowledge the effort of the CESR staff in providing us with
excellent luminosity and running conditions.
J.P.A., J.R.P., and I.P.J.S. thank                                           
the NYI program of the NSF, 
M.S. thanks the PFF program of the NSF,
G.E. thanks the Heisenberg Foundation, 
%
%
K.K.G., M.S., H.N.N., T.S., and H.Y. thank the
OJI program of DOE, 
J.R.P., K.H., M.S. and V.S. thank the A.P. Sloan Foundation,
R.W. thanks the 
Alexander von Humboldt Stiftung, 
M.S. thanks Research Corporation, 
and S.D. thanks the Swiss National Science Foundation 
for support.
This work was supported by the National Science Foundation, the
U.S. Department of Energy, and the Natural Sciences and Engineering Research 
Council of Canada.
\nobreak

\begin{table}
\caption{Results of the $\Lambda^+_ce^-X$, 
$\Lambda^+_c\overline{p}e^-\overline{\nu}_e$, and
$\Lambda^+_c\overline{p}X$ analyses.  The ``Data ON'' row shows results from 
fits to the on resonance data sample, while the ``Scaled OFF'' row shows the 
results of the fits to the nearby continuum data scaled for luminosity and 
cross section. }  
\label{tab1}
\begin{tabular}{ccccc}
TYPE & $\Lambda^+_ce^-X$ &$\Lambda^+_c\overline{p}e^-\overline{\nu}_e$
& $\Lambda^+_c\overline{p}X$ \\ \hline
Data ON 	& $176\pm 41$ & $20\pm10$ &$ 2501\pm121$\\ 
Scaled OFF 	& $9\pm 22$ & $6\pm7$ &$440\pm105$\\ 
Fakes   	& $10\pm 4$ ($e$ ) & $2\pm1$($e$ and $\overline{p}$ ) 
&$32^{+6}_{-15}$($\overline{p}$'s)\\
MC pred. uncorr. & $95\pm 7\pm 33$ & $11\pm6$ &$58\pm6\pm58$\\ \hline
Bkgd. sub. 	& $62\pm 47\pm 34$ & $1\pm12\pm6$ &$1971\pm160
^{+59}_{-60}$\\
Efficiencies  	& $0.239\pm 0.005\pm0.011$ &$0.094\pm0.003\pm0.003$ &
$0.257\pm0.003\pm0.010$  \\  \hline
YIELD & $259\pm 196\pm 143$ &$ 11\pm132\pm82$ &$7669\pm623\pm385$
\end{tabular}
\end{table}


\vfill
\eject
\begin{figure}
\vspace{7.5 cm }
\includegraphics{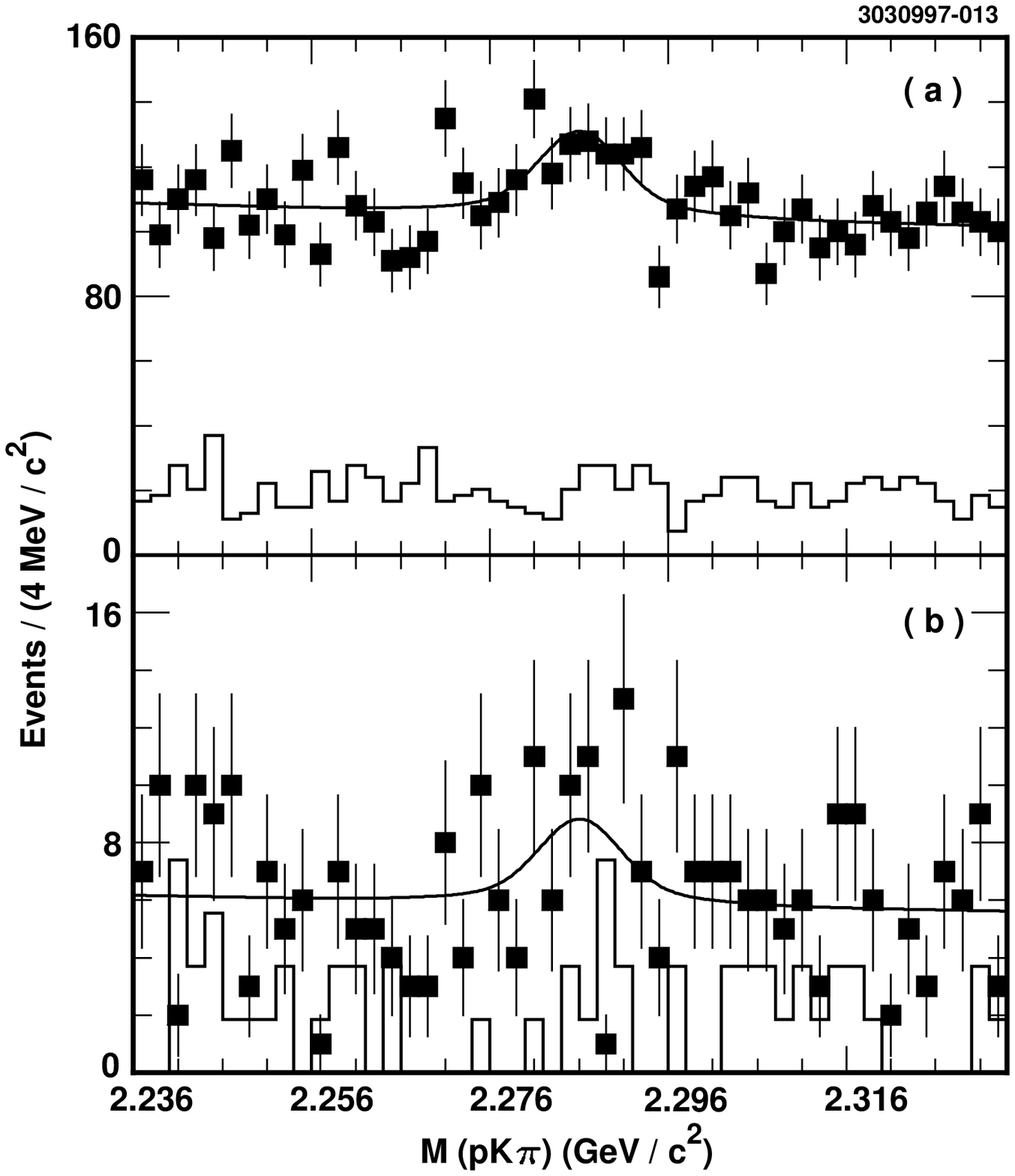} 
\caption{The fit (line) to the $pK^-\pi^+$ invariant mass spectrum from 
on resonance data events (points with error bars) and the scaled off 
resonance data (histogram) for the 
(a) $\overline{B}\rightarrow\Lambda^+_ce^-X$
and (b) $B^-\rightarrow\Lambda^+_c\overline{p}e^-\overline{\nu}_e$
 analyses.}
\label{fig1}
\end{figure}

\begin{figure}
\vspace{7.5 cm}
\includegraphics{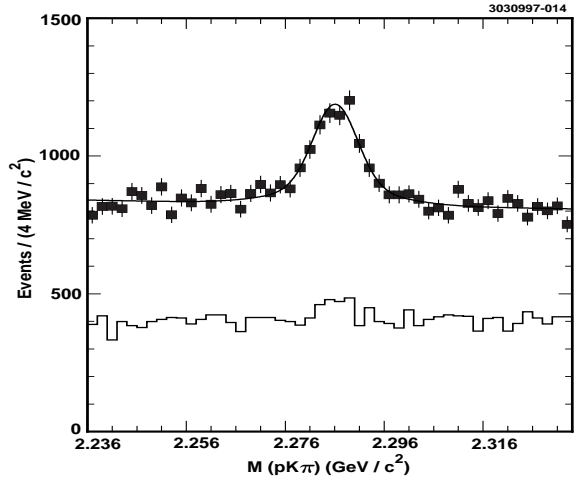}
\caption{The fit (line) to the $pK^-\pi^+$ invariant mass spectrum from 
on resonance data events (points with error bars) and the scaled off 
resonance data (histogram) for the 
$\overline{B}\rightarrow\Lambda^+_c\overline{p}X$ analysis.}
\label{fig2}
\end{figure}


\begin {thebibliography}{10}
\bibitem{ref3}G. Crawford {\it et al.}, Phys. Rev. {\bf D45}, 752 (1992).
\bibitem{ref5}Review of Particle Properties, Phys. Rev. {\bf D54}, 1 (1996).
\bibitem{ref4}R. Ammar {\it et al.}, Phys. Rev. {\bf D55}, 13 (1997).
\bibitem{ref20}M. Procario {\it et al.}, Phys. Rev. Lett. {\bf 73}, 
1472 (1994).
\bibitem{ref1}I. Bigi {\it et al.}, Phys. Lett. 
{\bf B323}, 408 (1994).
\bibitem{ref2}I. Dunietz, P. Cooper, A. Falk, and M. Wise, Phys. Rev. Lett. 
{\bf 73}, 1075 (1994).
\bibitem{ref11}Y. Kubota {\it et al.}, Nucl. Instrum. Methods Phys. Res., Sec. A {\bf 320}, 66 (1992). 
\bibitem{ref13}G.C. Fox and S. Wolfram, Phys. Rev. Lett. {\bf 41}, 1581 (1978).
\bibitem{ref9}H. Albrecht {\it et al.}, Phys. Lett. {\bf B249}, 359 (1990).
\end {thebibliography}
\end{document}